\documentclass[twocolumn]{aastex61}

\usepackage{tcolorbox}
\usepackage{savesym}
\usepackage{soul}

\usepackage{amsmath}
\usepackage{graphicx}
\savesymbol{tablenum}
\usepackage{etoolbox}
\usepackage{verbatim}
\usepackage{siunitx}
\restoresymbol{SIX}{tablenum}
\usepackage[caption=false]{subfig}
\definecolor{orcidlogocol}{HTML}{A6CE39}

\shorttitle{Confirmation of the Stellar Binary Microlensing Event, Macho 97-BLG-28}
\shortauthors{Blackman et al.}

\begin{document}

\title{Confirmation of the Stellar Binary Microlensing Event, Macho 97-BLG-28} 

\author[0000-0001-5860-1157]{J.W. Blackman}
\affil{School of Natural Sciences, University of Tasmania,
Private Bag 37 Hobart, Tasmania 7001 Australia}
\affil{Sorbonne Universit\'es, UPMC Universit\'e Paris 6 et CNRS, 
UMR 7095, Institut d'Astrophysique de Paris, 98 bis bd Arago,
75014 Paris, France}

\author{J.-P. Beaulieu}
\affil{School of Natural Sciences, University of Tasmania,
Private Bag 37 Hobart, Tasmania 7001 Australia}
\affil{Sorbonne Universit\'es, UPMC Universit\'e Paris 6 et CNRS, 
UMR 7095, Institut d'Astrophysique de Paris, 98 bis bd Arago,
75014 Paris, France}

\author{A.A. Cole}
\affil{School of Natural Sciences, University of Tasmania,
Private Bag 37 Hobart, Tasmania 7001 Australia}

\author{A. Vandorou}
\affil{School of Natural Sciences, University of Tasmania,
Private Bag 37 Hobart, Tasmania 7001 Australia}

\author{N. Koshimoto}
\affil{Department of Earth and Space Science, Graduate School of Science, Osaka University, 1-1 Machikaneyama, Toyonaka, Osaka 560-0043, Japan}

\author{E. Bachelet}
\affil{Las Cumbres Observatory, 6740 Cortona Drive, Suite 102, Goleta, CA 93117 USA}

\author{V. Batista}
\affil{Sorbonne Universit\'es, UPMC Universit\'e Paris 6 et CNRS, 
UMR 7095, Institut d'Astrophysique de Paris, 98 bis bd Arago,
75014 Paris, France}

\author{A. Bhattacharya}
\affil{Laboratory for Exoplanets and Stellar Astrophysics, NASA/Goddard Space Flight Center, Greenbelt, MD 20771, USA}

\author{D.P. Bennett}
\affil{Laboratory for Exoplanets and Stellar Astrophysics, NASA/Goddard Space Flight Center, Greenbelt, MD 20771, USA}


\begin{abstract}
The high-magnification microlensing event MACHO-97-BLG-28 was previously determined to be a binary system composed either of two M dwarfs, or an M dwarf and a brown dwarf. We present a revised light-curve model using additional data from the Mt. Stromlo 74" telescope, model estimates of stellar limb darkening and fitting the blend separately for each telescope and passband. We find a lensing system with a larger mass ratio, $q = 0.28 \pm 0.01$, and smaller projected separation, $s = 0.61 \pm 0.01 $, than that presented in the original study. We revise the estimate of the lens-source relative proper motion to $\mu_{rel}=2.8 \pm 0.5 \; \mathrm{mas\: yr^{-1}}$, which indicates that 16.07 years after the event maximum the lens and source should have separated by $46 \pm 8 \;\si{mas}$. We revise the source star radius using more recent reddening maps and angular diameter-color relations to $R_*=(10.3 \pm 1.9) R_\odot$. K and J-band adaptive optics images of the field taken at this epoch using the NIRC2 imager on the Keck telescope show that the source and lens are still blended, consistent with our light-curve model. With no statistically significant excess flux detection we constrain the mass, $M_L= 0.24^{+0.28}_{-0.12}M_\odot$, and distance, $D_L = 7.0 \pm 1.0 \,\si{kpc}$, of the lensing system. This supports the interpretation of this event being a stellar binary in the galactic bulge. This lens mass gives a companion mass of $M=0.07^{+0.08}_{-0.04}M_\odot$, close to the boundary between being a star and a brown dwarf.
\end{abstract}


\received{June 11, 2019}
\revised{January 5, 2020}
\accepted{January 6, 2020}
\submitjournal{ApJ}


\section{Introduction} \label{sec:intro}

Gravitational microlensing can be used to detect exoplanets beyond the snow line \citep{Mao1991,Gould1992, Bennett1996}. Unlike radial velocity and transit detection methods, microlensing does not depend on the luminosity of the host star. This means it is possible to effectively detect planets orbiting the M-dwarfs in the Galactic bulge. Properties of microlensing systems such as the lens-source angular separation and mass ratio can be determined by modelling the observed photometric lightcurve. In the absence of second order effects such as parallax or the influence of the finite-size of the source \citep{Beaulieu2006,Gaudi2008,Han2013}, it is often difficult to robustly determine the physical parameters of these systems.\\
\indent One way to constrain lens properties is by obtaining high angular resolution observations of the source in the years after the event. By this time the relative lens-source proper motion, typically of the order of 4-8 mas $\si{yr^{-1}}$, may be such that the source and lens can be resolved \citep{Bennett2015, Batista2015}. Even if they are still blended, the difference between the modelled baseline source flux and the high angular resolution measurement can be used to give an estimate of the lens brightness. These follow-up data can be obtained at large ground-based telescopes equipped with adaptive optics (AO) systems such as Subaru \citep{Fukui2015} and Keck \citep{Beaulieu2016, Beaulieu2018, Batista2013, Sumi2016}. Even if they cannot be resolved, distortion in the PSF due to the unresolved source-lens may be able to be detected. This was achieved for the first time for OGLE-2012-BLG-0950Lb in \cite{Bhattacharya2018} as part of the development of a mass-measurment method to be employed with the WFIRST space telescope (the \textit{Wide Field InfraRed Survey Telescope}, \cite{Spergel2015}).\\
\indent In this paper we revisit the microlensing event MACHO-97-BLG-28 \citep[hereafter A1999]{Albrow1999}, following the approach outlined in \cite{Batista2014,Batista2015} and \cite{Beaulieu2016}. This event occurred prior to the first microlensing exoplanet detection, early in the history of microlensing. It was notable for being the first time a central caustic cusp crossing was observed, and the first time limb darkening (LD) coefficients were determined by microlensing. A1999 determined the lensing systems of MACHO-97-BLG-28 to be stellar binary with mass ratio $q = 0.234$ and projected separation (in units of Einstein Rings Radius, $\theta_E$) $s = 0.686$. Two possible solutions were presented from their light curve model: either the lens lies in the galactic bulge and is likely to be a stellar M-dwarf binary with a separation of 1-2 AU, or the lens lies closer in the galactic disk. In this case one or both of the lens objects would be brown dwarfs. A subsequent compilation study \citep{Alcock2000} remodelled the event and found results consistent with this interpretation.\\
\indent The way that microlensing events are modelled has evolved since the event reached maximum magnification on 18 October 1997. Here we remodel this event using limb darkening (LD) coefficients from \cite{Claret2000}, instead of deriving the coefficients from the microlensing model. The fitted LD coefficients from A1999 are in disagreement with that from   \cite{Claret2000}, and this impacts on the derived source radius. Secondly, rather than adopting a global blend, we model the blend for each telescope and each band individually. We refit the light curve of MACHO-97-BLG-28 using the original data set with these modifications. We model the event using the open-source microlensing modelling package, pyLIMA \citep{Bachelet2017}, both with and without additional unpublished data from the Mt. Stromlo (MSO) 74'' telescope. Results of these models are presented in Section 2.\\ 
\indent Finally, in an effort to obtain an accurate mass measurment of the lensing system, we observe this object with Keck Adaptive Optics in an effort to resolve the predicted bright sub-arcsecond blend. 16.07 years after the event's peak, high angular resolution images were obtained of the source and blend in J and K-band in July 2013. The relative lens-source proper motion, $\mu_{rel}$, and the relative faintness of the lens compared with the source is such that both objects are still blended in the 60-80 mas seeing of our Keck images. A review of the current status of using adaptive optics observations in this manner can be found in \cite{Beaulieu2018a}. The results of these Keck observations are presented in Section 3. 
\begin{figure*}[ht!]
\centering
\includegraphics[width=\textwidth,height=\textheight,keepaspectratio]{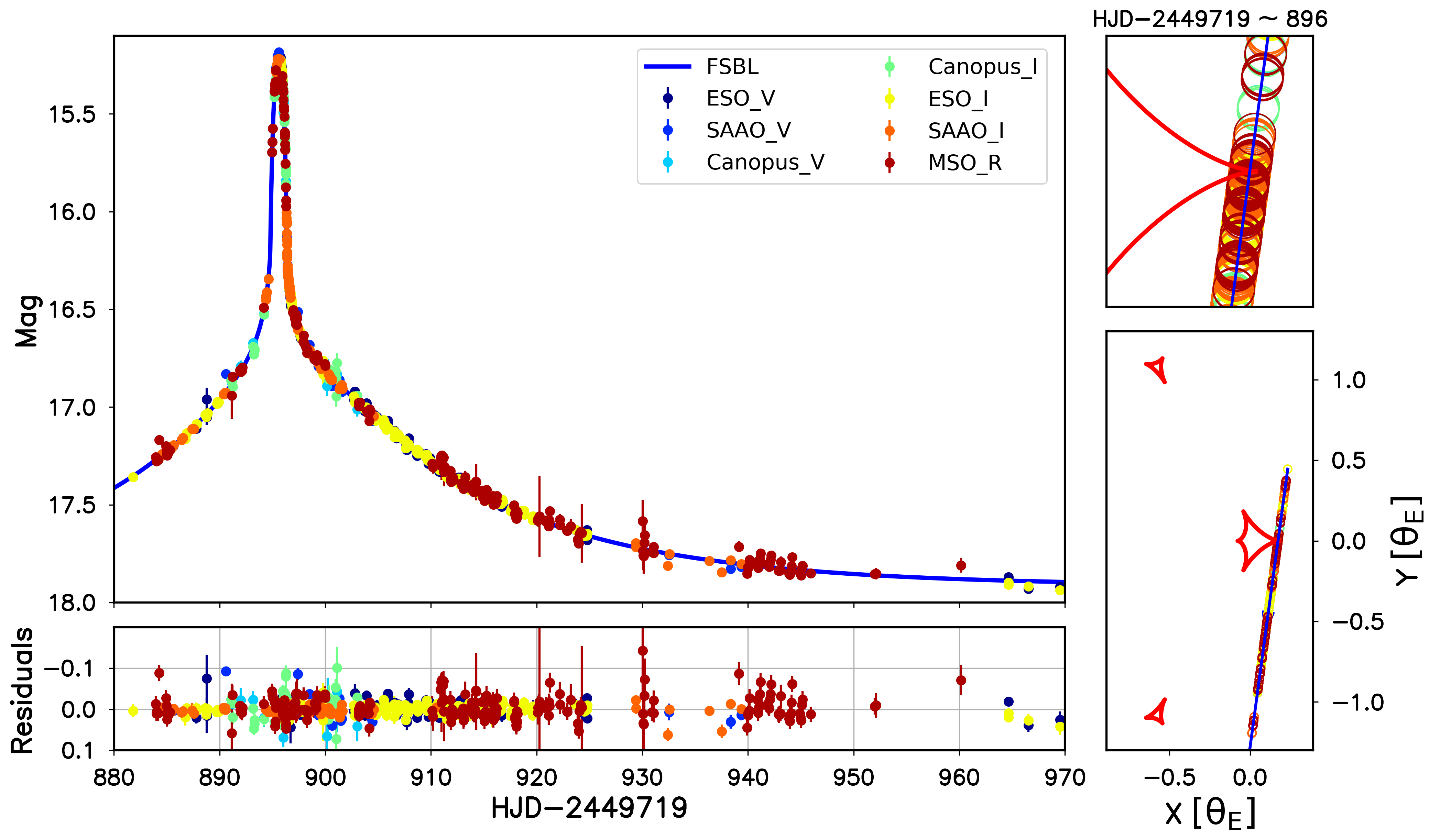}
\caption{The light curve for MACHO-97-BLG-28 with the best fit binary model (MOD2 Rescaled). V-magnitudes are shown on the y-axis. The insert shows the caustic geometry for the MOD2 rescaled model. The blue line indicates the source trajectory and the size of the circles crossing the central caustic indicates the source size. A zoom in on the peak for each of the four models can be found in Fig. \ref{fig:zoom} \label{fig:lc}}
\end{figure*}
\section{Modelling the MACHO-97-BLG-28 Light Curve with \textit{\MakeLowercase{py}LIMA}} \label{sec:model}
The microlensing event MACHO-97-BLG-28 is located at $\alpha = 18^h00^m33^s.8$ and $\delta = -28^{\circ}01'10''$ and was first observed following an alert from the Microlensing Planet Search (MPS) on May 29, 1997. Dense monitoring by The PLANET (Probing Lensing Anomalies NETwork) collaboration followed the high magnification increase in brightness on June 14. Observations continued for six weeks with additional baseline data points obtained in March 1998. The full V and I-band data set features good coverage of the event-timescale, with the exception of the baseline pre-magnification, and the sharp increase in brightness at HJD -- 2449719 = 894.8 (See Fig. \ref{fig:lc}).\\
\indent A1999 defined a high quality subset of their data under the condition that $I \lesssim 14.7$ and $V \lesssim 17$ with a $FWHM < 2.2\arcsec$. These are the data  we use in our modelling. This includes 431 I-band images (247 from the Dutch/ESO 0.92m at La Silla; 130 from the SAAO 1m; and 54 from the Canopus 1m in Tasmania) and 155 V -band (98 La Silla, 41 SAAO, 16 Canopus). 267 R-band images were also obtained by the MSO 74". These R-band data cover a broad range of the event timescale and are particularly significant during the upswing towards the peak following the first caustic crossing. Unused in the original study, we augment our models with these extra data points.\\
\subsection{Error-bar rescaling}\label{sec:rescale}

When modelling microlensing light curves, error bars are frequently rescaled in order to compensate for low-level systematics and general underestimation of uncertainties. Typically they are rescaled as: \begin{equation}\sigma'= k\sqrt{\sigma+e_{min}^2}\end{equation} where $k$ is a linear scaling factor and $e_{min}$ a minimum error added in quadrature. 
As in \cite{Bachelet2018} we avoid using the traditional metric of forcing $\chi^2/dof = 1$ as it is is only relevant for linear models \citep{Andrae2010}. Instead we run Kolmogorov-Smirnov and Anderson-Darling residual normality tests in order to optimise error bar scaling. These tests were applied to the $ESO_I$, $SAAO_I$ and $MSO_R$ datasets only, as normality is difficult determine for datasets with $n<100$. Rescaling was performed when the p-value associated to the test was less than 1\% (i.e. the test failed). With no rescaling $ESO_I$ passed the Kolmogorov-Smirnov test but failed the Anderson-Darling test, while for $SAAO_I$ the inverse was true. $MSO_R$ failed both tests. We adopt an $e_{min} = 0.005$ for the ESO and SAAO datasets, and a $e_{min}$ = 0.01 for the lower precision Canopus and MSO data, the choice of which is based on the quality of data obtained these sites. The $e_{min}$ term is particularly relevant to these data as there is very little coverage of the baseline and it only becomes significant when the event is bright. The multiplicative $k$ factor was introduced where the $e_{min}$ was not sufficient to make the normality tests converge, and to compensate for the underestimated error of the DoPHOT photometry. The scaling factor for $SAAO_I$ was introduced in the other datasets with fewer data points ($SAAO_V$ and $Canopus_{V, I}$) for consistency. In the case of $MSO_R$ we adopt a $k=1.8$. A Shapiro-Wilk normality test was also performed, but no choice of scaling factors could force $ESO_I$ and $MSO_R$ to pass. The choice of rescaling parameters can be seen in Table \ref{tab:rescale}.

\begin{deluxetable}{c|ccc}
\tablecaption{Uncertainty rescaling coefficients used in our modelling\label{tab:rescale}}
\tablehead{\colhead{Name} & \colhead{$N_{data}$} & \colhead{k} & \colhead{$e_{min}$}}
\startdata
$ESO_V$       &  98   & 1.0  & 0.005  \\
$SAAO_V$                   & 41 & 1.2 & 0.005 \\
$Canopus_V$                  & 16  & 1.2 & 0.01  \\
$ESO_I$                      & 247   & 1.0 & 0.005  \\                                                             
$SAAO_I$                      & 130             & 1.2   &  0.005   \\
$Canopus_I$       & 54  & 1.2 & 0.01\\
$MSO_R$ & 267  & 1.8 & 0.01 \\
\enddata
\end{deluxetable}

\subsection{Modelling with pyLIMA}

\begin{figure*}
\centering
     \subfloat[MOD1\label{MOD1}]{%
       \includegraphics[width=0.48\textwidth]{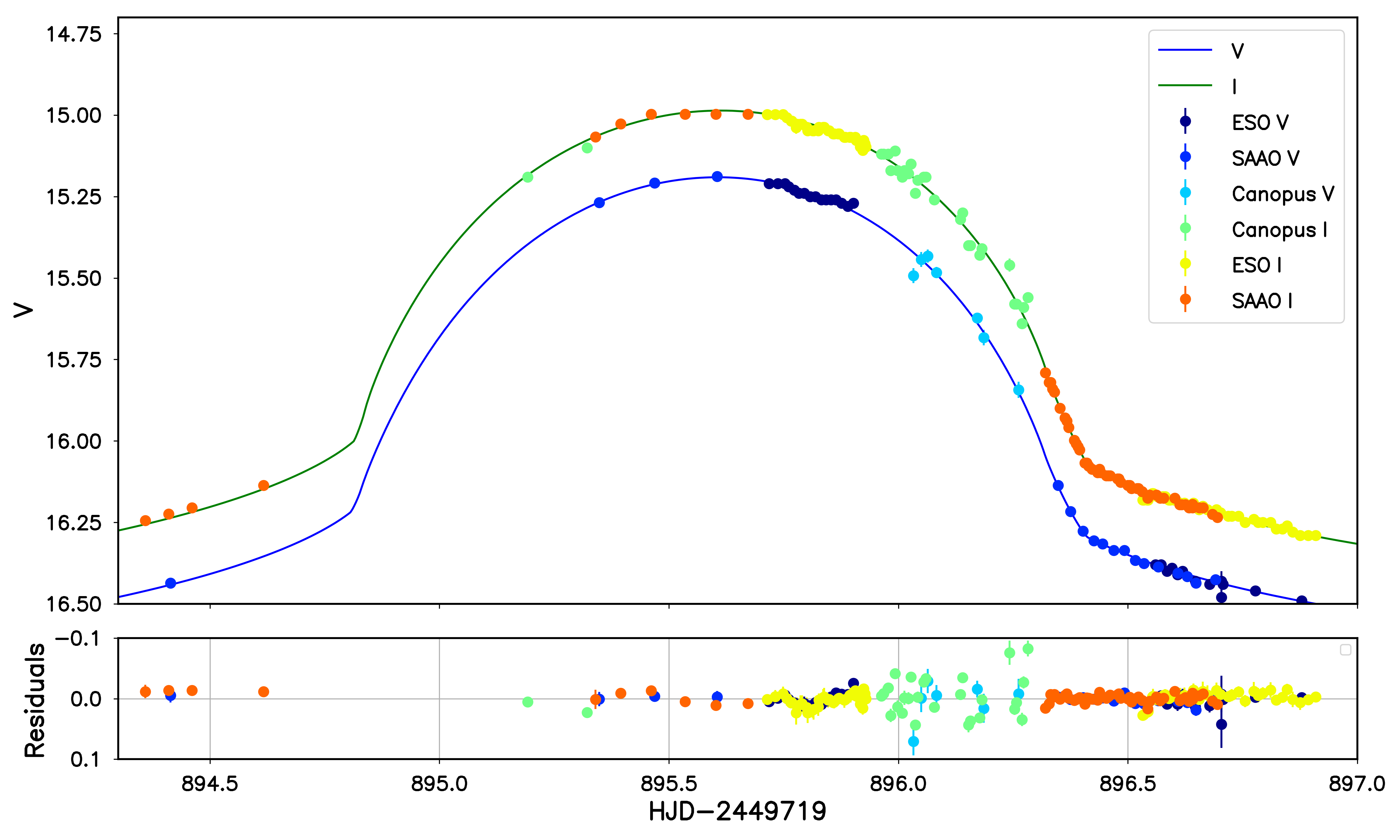}
     }
     \hfill
     \subfloat[MOD1 Rescaled\label{MOD1R}]{%
       \includegraphics[width=0.48\textwidth]{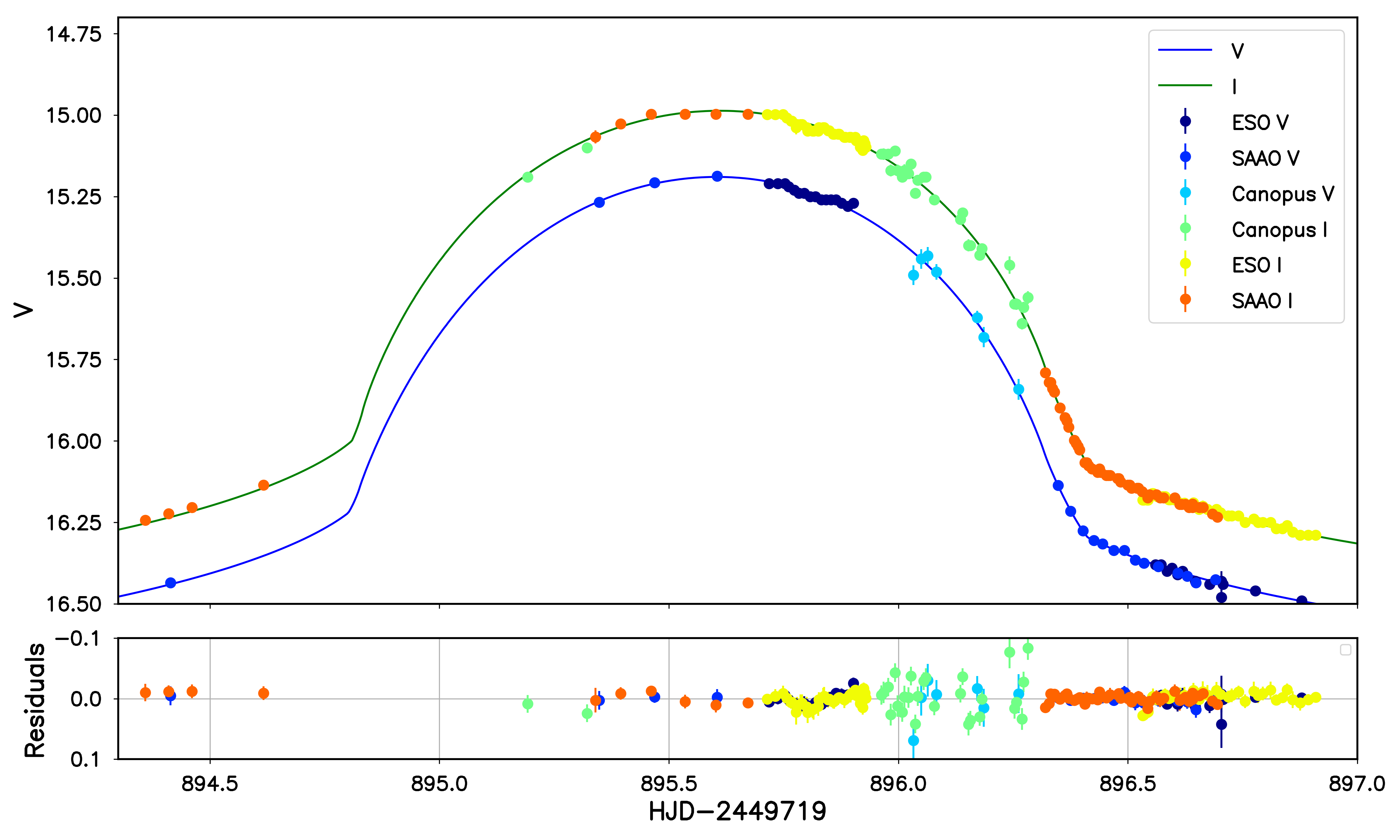}
     }
     \hfill
     \subfloat[MOD2\label{MOD2}]{%
       \includegraphics[width=0.48\textwidth]{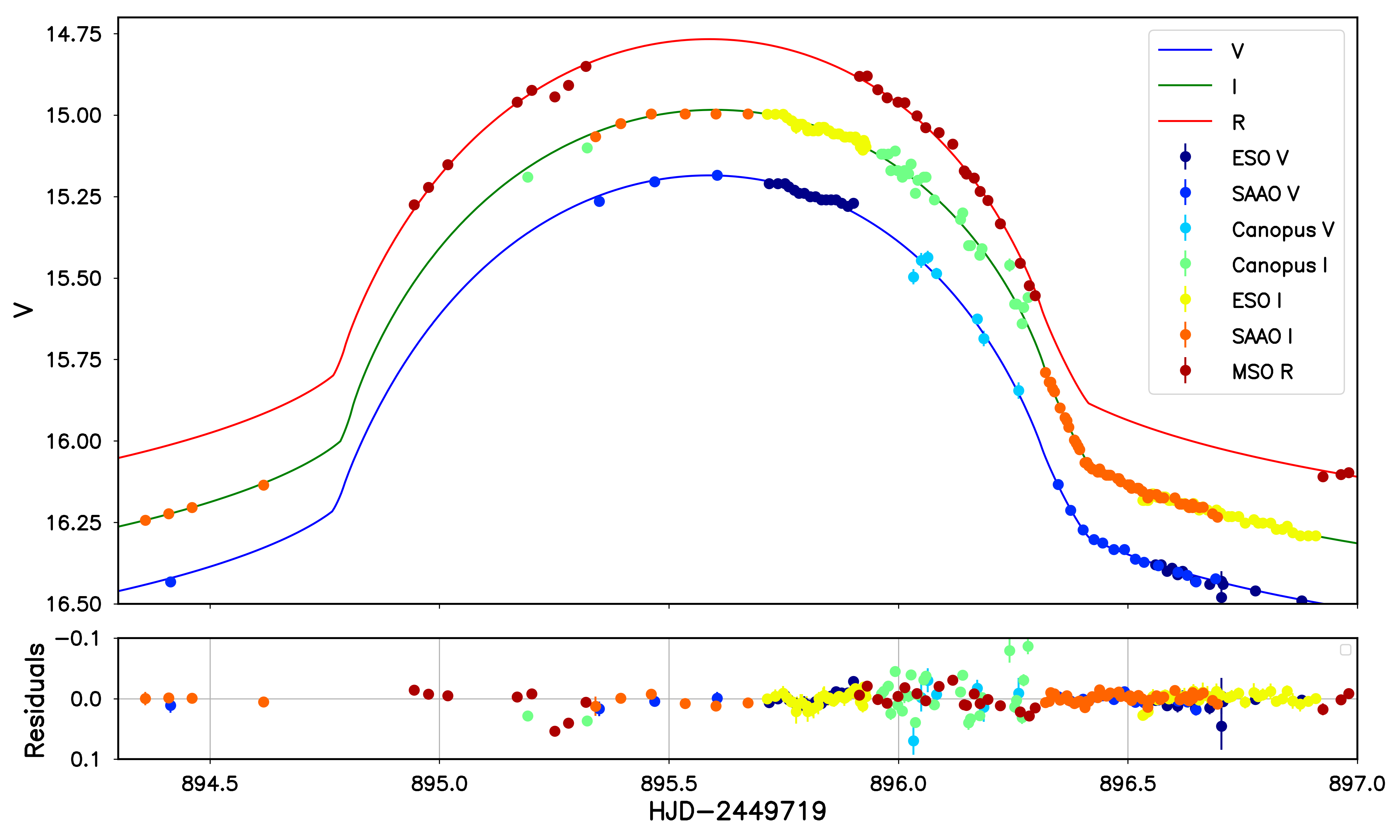}
     }
     \hfill
     \subfloat[MOD2 Rescaled\label{MOD2R}]{%
       \includegraphics[width=0.48\textwidth]{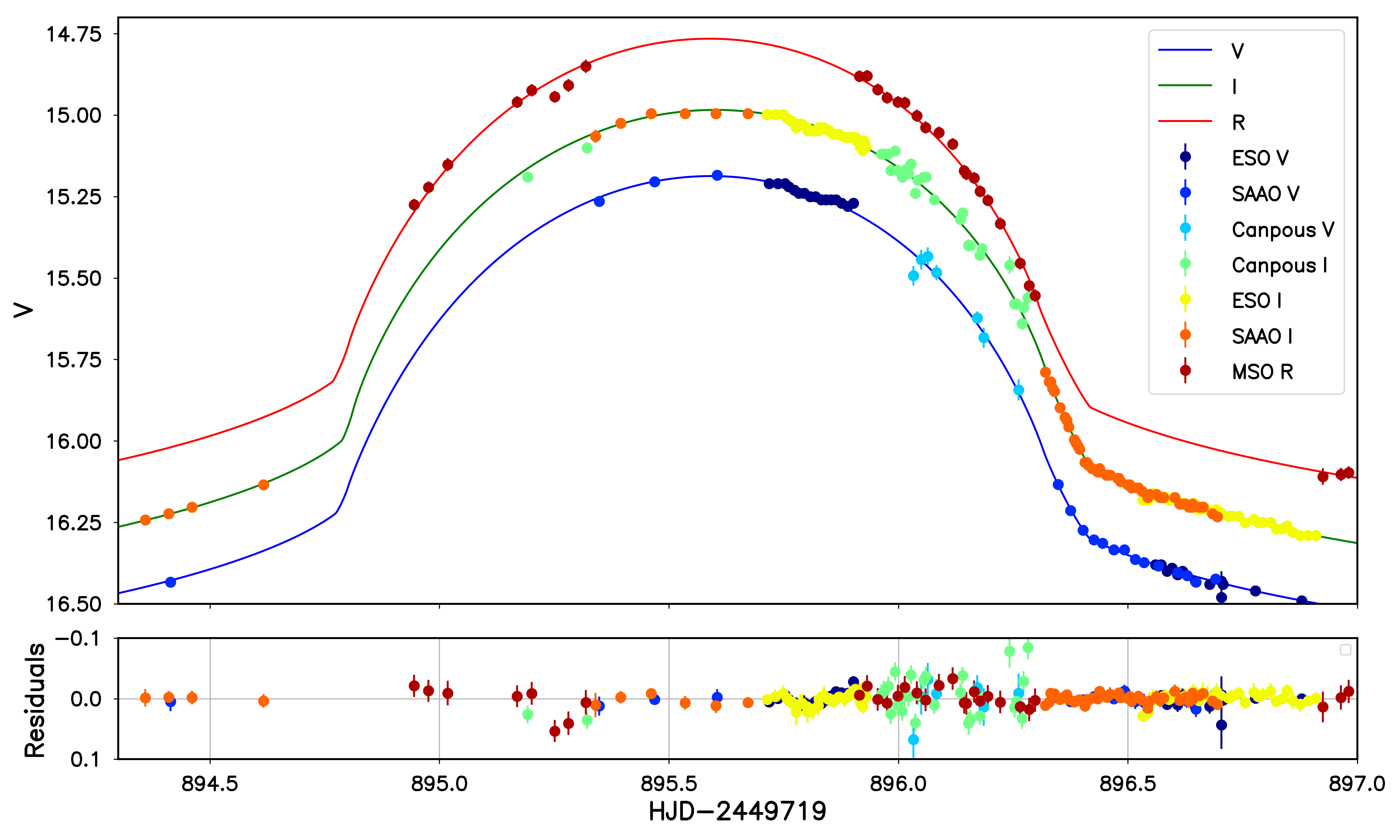}
     }
    \caption{Zoom on the event peak for all four models, with the light curves plotted for each band (V, I and R). The I and R light curves have been offset by 0.2 and 0.4 magnitudes, respectively.\label{fig:zoom}}
\end{figure*}

We use the Python microlensing modelling package pyLIMA to model the event light curve. pyLIMA is the first open-source package designed specifically for microlensing \citep{Bachelet2017}, and is available on the Github platform\footnote{https://github.com/ebachelet/pyLIMA}. In the first instance we perform two fits, MOD1 and MOD2, the results of which are shown in Table \ref{tab:result}. MOD1 features the exact same high quality subset of data used in the original study, with no rescaling. MOD2 includes the addition of the 267 MSO R-band data points. We also add two further light-curve models, MOD1 Rescaled and MOD2 Rescaled, also shown in Table \ref{tab:result}. In these two instances the uncertainties have been scaled according to the coefficients shown in Table \ref{tab:rescale}.\\
\indent To speed up processing we perform these fits in two stages. First we ignore limb darkening and constrain the grid search with a Uniform Source, Binary Lens (USBL) model using the differential evolution algorithm. Differential Evolution is a global optimizer first presented in \cite{Storn1997}. Surface brightness profiles are then determined for each band (I, V and R; \cite{Southworth2015}) by interpolating from the limb darkening coefficients presented in \cite{Claret2000}. We base our interpolation around the spectral type of the source star determined in section \ref{sec:sourcestar}.  With an intrinsic color of $(V-I)_0= 1.07 \pm 0.15$ we estimate $T_{eff} = 4600K$ and adopt $log g = 2.85$, $log [M/H] = 0.0$ and  $V_{micro} = 2  \: \mathrm{km/s}$. The limb-darkening coefficients that result are $u_I = 0.620$, $u_V = 0.799$ and $u_R = 0.718$. Once the most likely parameter values have been obtained from the USBL fit, we perform the more computationally intensive Finite Source Binary Lens (FSBL) starting with these initial guesses. This reduced the FSBL computation time from 504 to 84 hours on a single-core i5-3210m processor. Details of the FSBL method used by PyLIMA can be found in \citep{Bachelet2017}. We then perform Monte-Carlo Markov Chain (MCMC) explorations for all four models, utilizing the emcee algorithm \citep{Foreman-Mackey2013} in pyLIMA.\\ 
\indent Unlike A1999 we fit one blend parameter for each telescope and each band. This gives us a total of six blend parameters instead of two, and seven in the case of our second model. The blend parameters from the A1999 LD2 fit are given as $f_I=0.97$ and $f_V=0.84$. This is defined as the ratio of the source and baseline flux, or the proportion of the light which is lens. These figures indicate a small blend. In our fits use the pyLIMA definition of the blend flux, $g_i$, 
\begin{equation}
\begin{aligned}
f_{i}(t)=f_{s, i}\left(A(t)+g_{i}\right) ; \quad g_{i}=\frac{f_{b, i}}{f_{s, i}}
\end{aligned}
\end{equation}
where A(t) is the magnification with time t. Compared with the small blend fractions from A1999 ($g_I = 0.031$, $g_V=0.19$) we find ESO blends in our MOD2 fit of $g_I=0.38\pm0.01$ and $g_V=0.59\pm0.02$. A significant blend is to be expected for a target in the densely populated bulge. For MOD1 we minimise over 21 parameters and find a $\chi^2_{min}$ = 1650 for 565 degrees of freedom (dof), less than A1999's preferred ``LD2'' model which was arrived at with a $\chi^2_{min}$ = 1913 over 567 degrees of freedom. 
For MOD2 we add the MSO data and minimise over 24 parameters. With a $\chi^{2}_{min} = 8298$ and 829 degrees of freedom, we find the addition of the MSO data does not improve the significance of our fit. The rescaled MOD1 and MOD2, with parameters shown in Table \ref{tab:rescale}, in an effort to compensate for the underestimate photometric uncertainty, shows a minor increase in significance with the additional of the Stromlo data resulting in the reduction of $\chi^{2}_{min}/DOF$ from 1.43 to 1.37. In all models, this is a binary event with a larger mass ratio but smaller separation than A1999. Thus, though all models are similar (see also Fig. \ref{fig:zoom}), and result in the same physical interpretation of the system, with the minor reduction in $\chi^{2}_{min}/DOF$, we choose MOD2 Rescaled as our preferred model. We hence find the event has a mass ratio $q = 0.28\pm0.01$ and separation $s = 0.61\pm0.01$. The finite source size is estimated to be $0.0251\pm0.0003$, smaller than that predicted in A1999.

\begin{deluxetable*}{c|cccccc}[!htb]
\tablecaption{Microlensing Parameters for the Microlensing Event MACHO-97-BLG-28\label{tab:result}}
\tablehead{
\colhead{Parameter} & \colhead{MOD1} & \colhead{MOD1 Rescaled} & \colhead{MOD2} & \colhead{MOD2 Rescaled} & \colhead{A1999} & \colhead{A2000}}
\startdata
$t_E$ [days]        & $30.5(3)$      & 31.0(6)     & 30.1(1)     & 30.7(5)        & 27.3 & 26.4\\
$t_0$                   & $896.31(1)$   & 896.30(2)   & 896.268(6)  & 896.29(2)     & 896.42 & 896.37\\
$u_0$                  & $0.177(4)$      & 0.171(6)    & 0.168(1)     & 0.172(5)     & 0.215 & 0.225\\ 
$s$ 		           & $0.626(7)$      & 0.61(1)    & 0.608(3)    & 0.61(1)        & 0.686 & 0.707\\                                                                      
$q$ 		           & $0.254(6)$      & 0.27(1)    & 0.288(3)         & 0.28(1)    & 0.234 & 0.210\\
$\alpha$[rad]       & $-1.702(2)$    & -1.705(3)   & -1.709(1)     & -1.705(3)    & -1.712 & -1.705\\
$\rho_*[10^{-3}]$ & $0.0248(3)$   & 0.0245(5)  & 0.0257(1)    & 0.0251(3) &  0.0286 & 0.0288(5)\\
DOF                     & 565       & 565        & 829       & 829      & 567 & 1404\\
$\chi^2_{min}$    & 1650      & 812        & 8298      & 1135    & 1913 & 2734.9\\
\enddata
\tablecomments{Presented here are four Finite Source Binary Lens (FSBL) fits for the MACHO-97-BLG-28 light curve, with the LD2 fit parameters from A1999 presented for comparison. These fits were achived by performing a global optimization with the Differential Evolution alogorithm in \textit{PyLIMA}, followed by Markov Chain exoplorations. The number in brackets notes the $68\%$ error in the last digit derived from the MCMC. MOD1 uses the same data set as A1999. MOD2 includes an extra 267 R-band data points from the MSO 74". MOD1 Rescaled and MOD2 Rescaled are the fits when the data are rescaled as per Table \ref{tab:rescale} and Section \ref{sec:rescale}. We define $u_0$ according to the centre of mass of the system, contrary to A1999, which defined it as centered on the binary midpoint (see Appendix A, \cite{Albrow2000}). The A1999 values above have been converted to the centre of mass parameterization.} 
\end{deluxetable*}

\subsection{Spectral Type and a New Estimate of the Source Star Radius} \label{sec:sourcestar}

With the determination of the $(V-I)$ source color we are able to arrive at a new estimation of the source star size and spectral type. A1999 obtained an estimate of the source star size by comparing the source  $(V-I)_0-I_0$ color with that of red clump stars in the Galactic Bulge, together with low-resolution spectroscopy obtained on the ESO Faint Object Spectrographic Camera on the ESO 3.6m in Chile. Using a function from \cite{StanekGarn98} to describe the I-band distribution of red clump stars, and isochrones from \cite{Bertelli1994}, A1999 estimated a source mean radius of $R_* = (15 \pm 2) R_\odot$. Here we use the OGLE extinction calculator\footnote{http://ogle.astrouw.edu.pl/cgi-ogle/getext.py}, which assumes the $E(J-K)$ measurements of \cite{Gonzalez2012}, to calculate the extinction towards the source using a natural neighbour interpolation of good points. We find reddening and an I-band extinction of $E(V-I) = 1.318 \pm 0.129$ and $A_I = 1.535$. The $(V-I)$ color of the source is determined using the our calibrated baseline magnitudes from our fit in section 2.2. With calibrated (and de-blended) source baseline magnitudes of $V = 18.45 \pm 0.05$ and $I= 16.06 \pm 0.05$ we find a $(V-I)=2.39 \pm 0.07$. Corrected for extinction we estimate the intrinsic color to be $(V-I)_0= 1.07 \pm 0.15$. Using \cite{Bessell1988} and \cite{Bessell1990}, we revise the estimate of the source from a K2 giant to a K0/K1 giant. We refer to the angular diameter--color relations presented in \cite{Adams2017}, which extends the relations presented in \cite{Boyajian2013, Boyajian2014} by using a sample of sample of dwarfs/subgiants and a sample of giant stars. We use the coefficients 
\begin{multline}
\log(2\theta_*(mas))=(0.535\pm0.027)+(0.490\pm0.046)(V-I)\\-(0.068\pm0.019)(V-I)^2-0.2I
\end{multline}
in order to determine the angular size of the source. This, combined with the extinction law from \cite{Gonzalez2012} and \cite{Nishiyama2009}, gives an angular radius $\theta_* = 6.0 \pm 1.1\: \mu as$ and a radius of $R_*=(10.3 \pm 1.9) R_\odot$, which is notably smaller than that predicted in A1999. Our angular source radius determination is however in good agreement with that determined in \cite{Alcock2000}, $\theta_*=6.58 \pm 0.90\: \mu as$.

\subsection{Angular Einstein Ring Radius}

When modelling this event our data was calibrated with the same SAAO calibration as A1999. As in A1999 this results in a modelled baseline source+blend flux of $I=15.66\pm 0.05$ and $V = 17.95\pm 0.05$. With an $ESO_V$ blend factor of $0.59 \pm 0.02$ and $ESO_I$ blend of $0.38 \pm 0.01$ (with the blend factor defined as the ratio of the blend and source flux) we find a de-blended source magnitude of:
\begin{equation}
\begin{aligned}
V_s = 18.45 \pm 0.05\\
I_s = 16.06 \pm 0.05
\end{aligned}
\end{equation}
Data coverage over the caustic crossing (see the top right panel in Fig. \ref{fig:lc})  indicates the presence of finite source effects, which means that we are able to determine the source radius crossing time ($t_*$) and estimate the relative source-lens proper motion. Combining our source-star radius calculation from \ref{sec:sourcestar} with our modelled finite size of the source, $\rho_*$, we can calculate the angular Einstein ring radius, \begin{equation}\theta_E=\theta_*/\rho \end{equation} We find an angular  Einstein ring radius of $\theta_E =0.24 \pm 0.04$ mas and a lens-source relative proper motion of $\mu_{rel}=2.8 \pm 0.5 \: \mathrm{mas\: yr^{-1}}$, the latter of which is smaller than the A1999 estimate of $\mu_{rel}= 4.09 \pm 0.55\: \mathrm{mas\: yr^{-1}}$. With the 16.07 years between observations we hence calculate the predicted lens-source separation at the time of our follow-up observations to be $46 \pm 8$ mas, which is of the order of the best FWHM photometry achievable from Keck.

\section{Adaptive Optics observations of MACHO-97-BLG-28}\label{sec:ao}
\begin{figure}[!h]
\centering
\includegraphics[width=0.95\columnwidth]{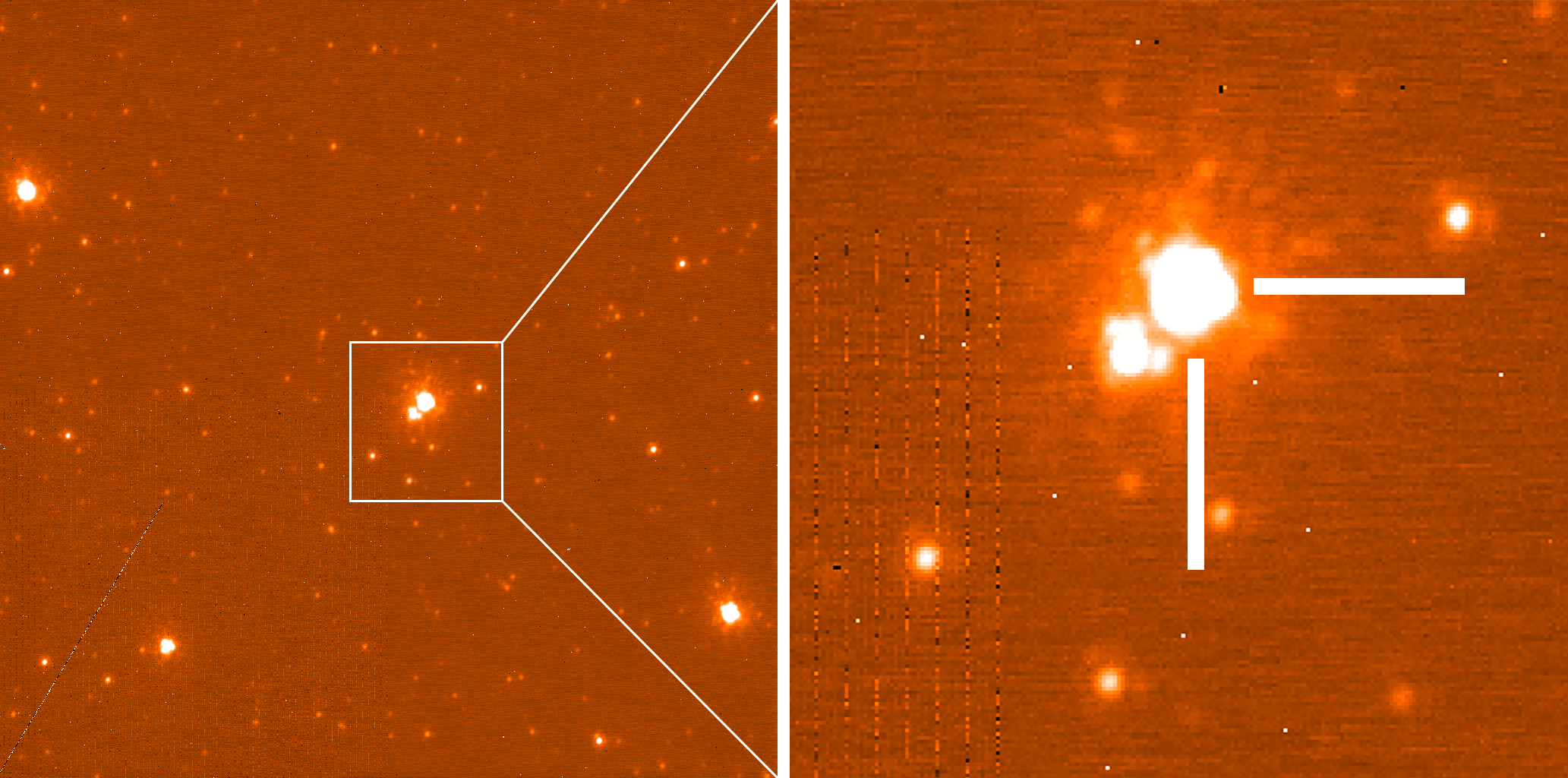}
\caption{Keck II K-band images of the source star field of the microlensing event MACHO-97-BLG-28, taken with the narrow camera on NIRC2. The left image is the full ten arcsecond Keck frame. The right image is a two arcsecond zoom centred on the source star.}\label{fig:keck}
\end{figure}
MACHO-97-BLG-28 was observed using the NIRC2 Adaptive Optics Imager with the Keck II telescope on Manua Kea on 13 July 2013 (HJD = 2456487.291), with the intent of resolving the source and lens -- or finding PSF distortions as in \cite{Bhattacharya2018} -- and hence being able to constrain the lens mass-distance relation. We use the narrow camera, which results in a plate scale of 0.01 arcsecond $\mathrm{pixel^{-1}}$. Ten images were obtained in J and K, each with an exposure time of 10 seconds and a dither of 0.7 arcseconds. Our images feature a median FWHM of 64 mas in K and 78 mas in J, which while greater than then predicted lens-source separation, still facilitates a measurements of the excess flux. We reduce these data following the procedure described in \cite{Beaulieu2016} and \cite{Batista2014}, beginning with standard dark current and flat-field corrections. We align a single Keck image with data from the VVV Survey \citep{Minniti2010}. As the precision of the dithering is greater than that of the pointing, astrometry was performed manually on the remaining images with the first used as a reference. A catalog of sources was generated from this reference image using \texttt{SExtractor} \citep{Bertin1996}, with it being used to realign each image in turn. The data were then stacked using \texttt{SWARP} \citep{Bertin2010} and the calibration constant calculated by cross-matching the K images with the VVV catalog. This process was repeated for the data in J-band. We find K and J-band magnitudes at the predicted position of the source to be:
\begin{equation}\label{eq:measure}(J,K)_{Keck}=(14.12, 12.92) \pm (0.07,0.06)\end{equation}
These magnitudes were calculated by comparing our photometry with the VVV magnitudes at the source position: $J_{VVV}= 13.92 \pm 0.02$ and $K_{VVV} = 12.76 \pm 0.02$.
\indent To determine if there is an excess flux detection, we compare this Keck measurement with the predicted K and J baseline magnitudes derived from our modelled unmagnified source flux, $(V,I)_{s,model} = (18.45, 16.06) \pm (0.05, 0.06)$. 
We perform a Monte Carlo simulation with $(V,I)_{b,model}$, $A_V$, $A_I$, $A_J$ and $A_K$ as parameters. Estimated distance modulus, age and metallicity are derived from isochrone models \citep{Bressan2012} as in \cite{Bennett2018}. We use isochrone models with $-0.3 < [M/H] < 0.3$ and $10\;Gyr < \textrm{age} < 13\;Gyr$, approximating the bulge, to derive $J$ and $K$. We find the (predicted) non-magnified source magnitudes to be 
\begin{equation} \label{eq:source}
\begin{aligned}
J = 14.18^{+0.20}_{-0.17}\\ 
K = 13.04^{+0.27}_{-0.22}
\end{aligned}
\end{equation}
The large error in these values are dominated by the uncertainty in metallicity and age. 
Comparing the measured source magnitudes (\ref{eq:measure}) with the predicted (\ref{eq:source}), we find no statistically significant excess flux detection.

\begin{figure}[t]
\centering
\includegraphics[width=0.95\columnwidth]{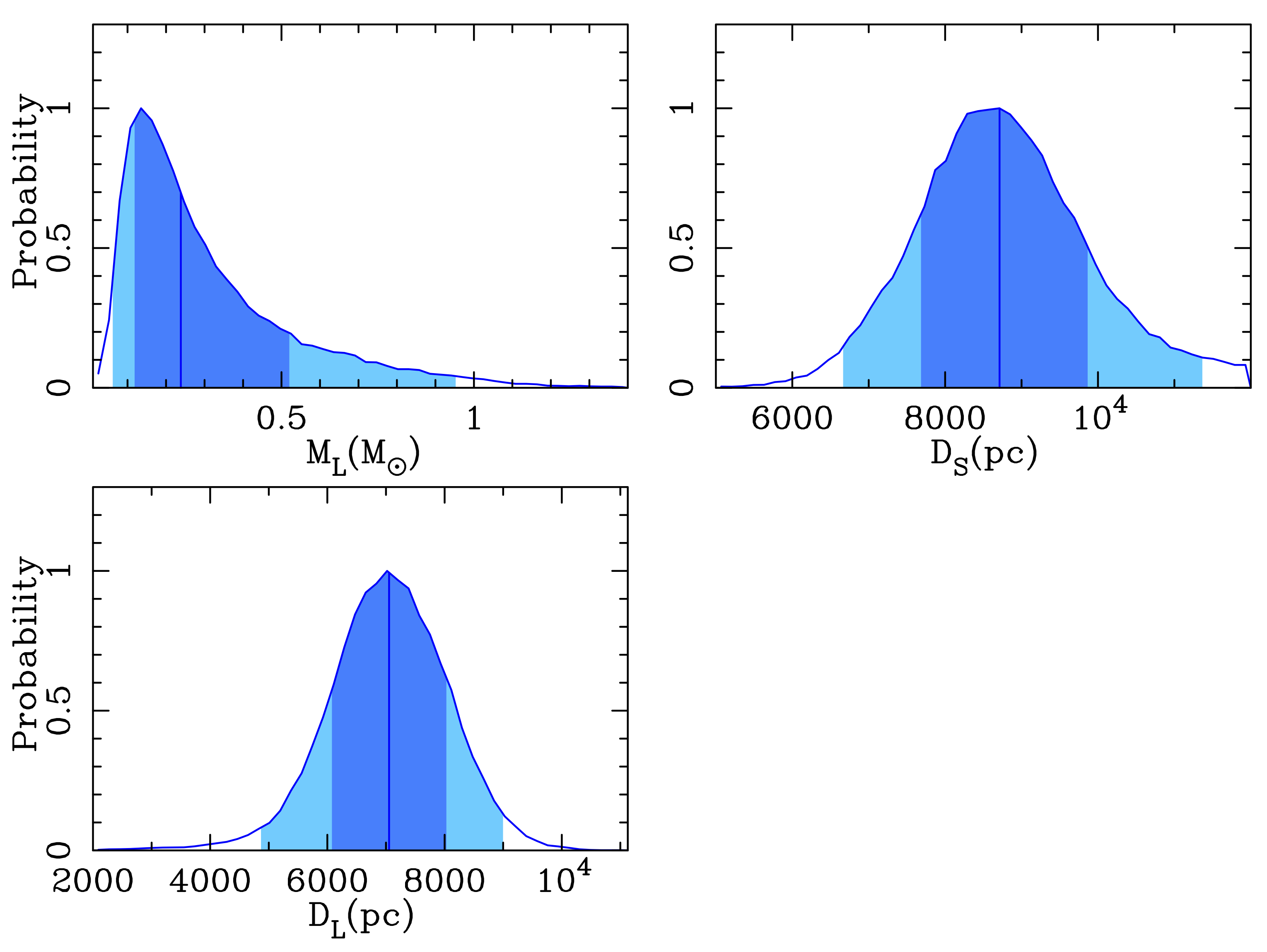}
\caption{Posterior probability distributions of the physical properties of the system determined using a galactic model \cite{Sumi2011}. The inner dark regions represent the 1$\sigma$ limits while the the light blue regions represent 2$\sigma$. Shown is $M_L$, the lens mass, $D_L$, the lens distance and $D_S$, the distance to the source.}
\label{fig:naokix}
\end{figure}
\begin{figure*}[!]
\epsscale{1}
\centering
\vspace{3mm}
\includegraphics{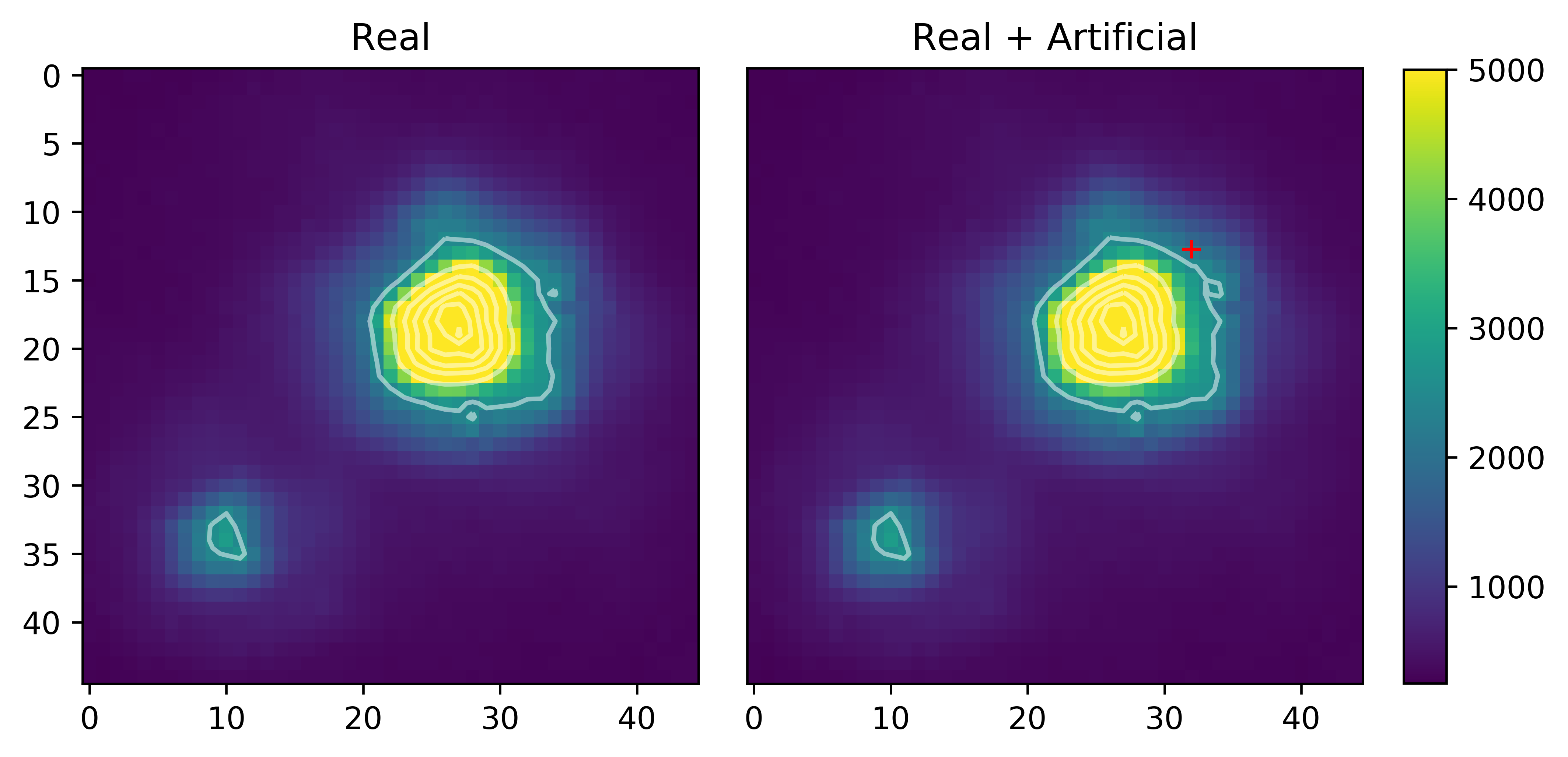}
\caption{Comparison of the Keck K-band images combined with \texttt{SWARP} (left), and the same image augmented with an artificial lens star at 63mas separation (right), generated by \texttt{DAOPHOT}. The axes in both images are in pixels, where each pixel represents 0.01 arcsec. The colour gradient represents the number of pixel counts. The position of the centroid of the artificial star is marked with a red cross. The separation and magnitude contrast are such that the source and lens could not be resolved if further AO observations of this event were made in 2019.}
\label{fig:aparna}
\end{figure*}
\subsection{Physical Parameters and Contemporary Observations}

To determine the physical parameters of the system we perform a Bayesian analysis using the galactic model from \cite{Sumi2011}. Posterior distributions of the source distance, $D_S$, lens mass, $M_L$ and lens distance, $D_L$ are presented in Fig. \ref{fig:naokix}. In this calculation we consider the mass ratio and separation as priors, and ignore the case where the lens is a remnant. With no excess flux detection, we weakly constrain the lens mass, $M_L= 0.24^{+0.28}_{-0.12}M_\odot$ and lens distance, $D_L = 7.0 \pm 1.0 \,\si{kpc}$. With our preferred MOD2 Rescaled mass ratio of $q=0.28\pm0.01$, this slightly favours the conclusion that the event is a stellar M-dwarf binary, though the other option proposed in A1999 of a M-dwarf and brown dwarf cannot be ruled out.\\
\indent To determine whether or not further AO observations might be worthwhile in 2019, we generate an artificial star on one of our K-band images using the \texttt{ADDSTAR} routine from \texttt{DAOPHOT} \citep{Stetson1987}. This star is placed 63mas from the centroid of the source -- the projected separation of the source-lens were that object to be observed in 2019. On inspection of the contour-plot images, seen in Fig. \ref{fig:aparna}, we find no significant difference between the real 2013 K-band image and the synthetic one. We conclude that the separation and magnitude contrast are such that the source and lens could not be resolved if further AO observations of this event were made in 2019.
\section{Discussion \& Conclusion} \label{sec:conclusion}

In this study we revisit the microlensing event MACHO 97-BLG-28 and confirm the interpretation of A1999 that it is a stellar binary. Following Fig. \ref{fig:naokix} we estimate a companion mass of $M= 0.07^{+0.08}_{-0.04}M_\odot$, right on the boundary between being a star and a brown dwarf. We use the open source microlensing code being developed for the next generation of microlensing studies, pyLIMA, and unused R-band data from Mt. Stromlo, to improve the robustness of the light-curve model. We adopt limb darkening parameters from stellar profile estimates determined subsequent to the original study and, as is standard in contemporary microlensing practise, fit blend parameters for each band and each telescope. Our refined parameters of this event find it to be a stellar binary with mass ratio $q = 0.28 \pm 0.01$, projected separation $s = 0.61 \pm 0.01$, and characteristic timescale $t_E=30.7\pm5\;\si{days}$, which is ${\sim}12.5\%$ longer than that predicted previously. Consistency is seen between different models, with only a modest improvement in $\chi^{2}_{min}/DOF$ from 1.43 to 1.37 following the introduction of the $MSO_R$ data. These new models are however shifted consistently from the A1999 model.\\ 
\indent This study is the first attempt to observe a stellar binary microlensing event with high resolution adaptive optics. It is also the first attempt to observe high resolution follow-up an event with a giant source star. Even though there was no statistically significant lens detection, the technique as presented here can be used on a) events with sufficient lens-source separation and proper motion that the source-lens can be resolved, and b) will be used in future using ELT-class telescopes such as GMT and TMT, which will have three times the resolution of Keck and hence be able to resolve them sooner. Further, this event has a projected separation of $\sim1.0AU$ and falls within the 0.5-10AU range for which microlensing is particularly sensitive \citep{Meyer2018}. An understanding of the planetary and stellar binary mass functions in this range is required if we are to better understand the brown dwarf desert, which is something that future adaptive optics studies such as this will be able to provide.\\ 


\indent We acknowledge the help of Dr. Peter Stetson in the use of DAOPHOT to generate artificial stars as in Fig. \ref{fig:aparna}. We acknowledge the support of the Astronomical Society of Australia for supporting the presentation of this work at the 22nd Microlensing Conference in Auckland, New Zealand, January 2018. This work was supported by the University of Tasmania through the UTAS Foundation and the endowed Warren Chair in Astronomy. Data presented in this work was obtained at the W. M. Keck Observatory from telescope time allocated to the National Aeronautics and Space Administration through the agencies scientific partnership with the California Institute of Technology and the University of California. 



\bibliographystyle{yahapj}
\bibliography{library}

\end{document}